# Computational genomic algorithms for microRNA-based diagnosis of lung cancer: the potential of machine learning

N. Garikipati

Huron High School, Ann Arbor, MI, USA

**Abstract**

The advent of large scale, high-throughput genomic screening has introduced a wide range of tests for diagnostic purposes. Prominent among them are tests using microRNA (miRNA) expression levels. Genomics and proteomics now provide expression levels of hundreds of miRNAs at a time. However, for actual diagnostic tools to become reality requires the simultaneous development of methods to interpret the large amounts of miRNA expression data that can be generated from a single patient sample. Because these data are in numeric form, quantitative methods must be developed. Statistics such as p-values and log fold change give some insight, but the diagnostic effectiveness of each miRNA test must first be evaluated. Here, the author has developed a traditional, sensitivity- and specificity-based algorithm, as well as a modern machine learning algorithm, and evaluated their diagnostic potential for lung cancer against a publicly available database. The findings suggest that the machine learning algorithm achieves higher accuracy (97% for cancerous and 73% for normal samples), in addition to providing confidence intervals that could provide valuable diagnostic support. The machine learning algorithm also has significant potential for expansion to more complex diagnoses of lung cancer sub-types, to other cancers, as well diseases beyond cancer. Both algorithms are available on the Github repo: https://github.com/neerja-g/machine-learning-miRNA.

1. **Introduction**

Lung Cancer is the leading cause of death among both male and female cancer patients, accounting for 25% of all cancer deaths [1]. There are over 430,000 Americans living today who have been diagnosed with lung cancer at some time. While the popular belief is that lung cancer is caused by the inhalation of carcinogens due to smoking, 20-30% of patients diagnosed with lung cancer have never smoked, leading researchers to believe that smoking is not the root cause of lung cancer. Additionally, not much is known about the genetic basis of this disease. Although significant advances have been made to treat and detect lung cancer, often, it is not detected until it is in an advanced stage, and is nearly untreatable.

microRNAs (miRNAs) are small RNA (Ribose Nucleic Acid) molecules that do not code proteins. Instead, they may "silence" the activity of certain RNAs, and regulate the actions of certain genes. Because of these functions, miRNA biomarkers have recently been used as a means to detect and predict diseases by using expression levels derived from tissue samples such as tumors. However, this method is often invasive, involves long recovery periods and can potentially lead to complications if performed incorrectly. In contrast, biomarkers derived from blood, urine, or saliva can serve the same purpose as other tissue biomarkers, while being significantly less invasive than their tissue. These advantages have made miRNA testing for lung cancer and other diseases more viable as a diagnostic approach. In parallel, there is now a growing number of studies that seek to develop techniques to interpret miRNA readouts. Because 100s or even 1000s of miRNAs can be tested for, the pROCessing of these results has fueled an increase in high throughput methods using bioengineering principles, prominent among them being lab-on-chip techniques. They have been complemented by quantitative methods in computational biology that use a range of statistical and mathematical algorithms. An early example was a study by Baldi & Long [2], who applied a Bayesian statistical analysis to DNA microarrays, which was found to be more effective than the simple t-test or fold methods. Lu et al. [3] conducted a study that used a novel, flow cytometric miRNA expression profiling technique to analyze miRNAs from samples, including cancerous ones. The expression levels reflected the development and differentiation of the tumors, and an overall downregulation of miRNAs was observed when compared to healthy tissue samples. Chin et al. [4] identified a single-nucleotide polymorphism (SNP) in an let-7 miRNA complementary site of the KRAS gene's 3' untranslated region (UTR) that increased risk for a variant called non-small cell lung cancer. Exosomal miRNA biomarkers from both cancer and non-cancer samples have been used as a benchmark to create a screening test that would indicate whether an individual had lung cancer [5]. MicroRNA expression levels have also been used in conjunction with CT scans as another method to diagnose lung cancer [6]. The study found that miRNA expression levels from healthy lung tissue showed key associations with clinical presentation, indicating the influence of a conducive microenvironment for developing tumors. Bianchi et al. [7] developed a diagnostic test derived from the detection of 34 miRNAs in serum that allowed them to identify still-early non-small cell lung carcinomas in asymptomatic patients who were high-risk individuals for lung cancer.

This study seeks to add to the body of research that develops computational genomic approaches to efficiently screen for lung cancer. It is based on expression levels of miRNAs in a large dataset obtained from the Gene Expression Omnibus (GEO) [8]. First, the sensitivity and specificity [9, 10] of the miRNAs in the dataset are examined to determine their suitability as diagnostic tests for lung cancer. Subsequently, a machine learning model is developed [11, 12] and trained on this dataset, as an alternate computational diagnostic tool. The two algorithms thus developed can be applied in future to individualized patient samples that have been tested for expression levels of the same array of miRNAs as in the dataset used here, with the goal of predicting the probability that a given patient sample is cancerous. They can also be extended to diagnosis of more complex sub-types of lung cancer, to other cancers and other diseases.

**2. Methods**

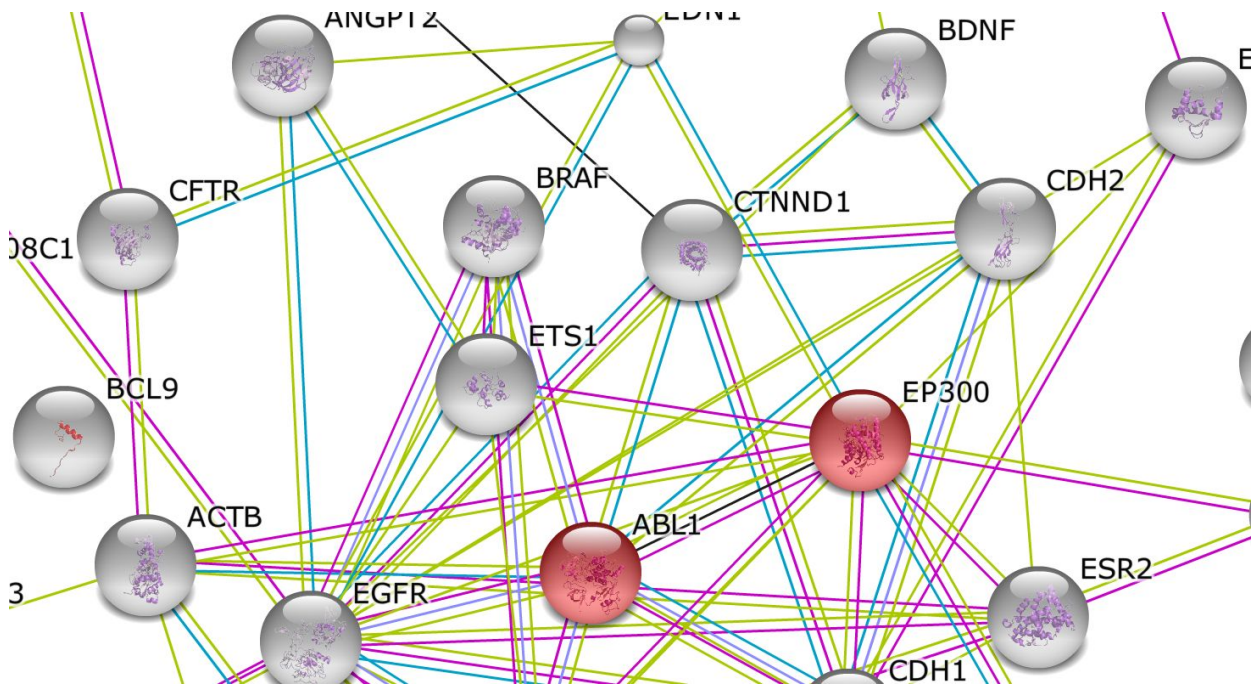

**Figure 1.** Apoptotic signaling pathways that could be downregulated by miRNAs in lung cancer.

The dataset used in this study was obtained from GEO [8], which provides RNA expression data for many different diseases. The specific dataset used here is GSE61741, and provided miRNA expression levels for 1049 patients, 73 of whom were known to have lung cancer, and 94 of whom were healthy. One possible role of miRNAs in the cancerous samples could be the downregulation of apoptotic signaling pathways (Figure 1). The other 882 patients had varying diseases, from glaucoma to sarcoidosis. Statistical analysis was performed on these two sample groups using GEO2R, GEO's integrated statistical analysis tool, and the p-value, adjusted p-value, t-value, B, and log fold change of all 848 miRNA samples were calculated. All algorithms were written and run as scripts in the freely available statistical software package, R.

## 2.1 Data analysis for sensitivity and specificity

One way to evaluate the usefulness of any test is to calculate its sensitivity and specificity. Sensitivity can be defined as the proportion of true positives in a test, that is, the proportion of tests that actually detected a disease state (test positives) relative to the total number of samples that are confirmed with the disease state (sample positives). Specificity is the proportion of tests that correctly detected a disease-free state (test negatives) relative to the total number of samples that were confirmed to be free of the disease (sample negatives). See Scholkopf et al. [9] and Gentleman et al. [10] for more extensive discussions of these statistical measures, and Davuluri et al. [13] or Mathe et al. [14] for applications in computational biology. In the example of Table 1, the sensitivity is $s^n = A/(A+C)$, and the specificity is $s^p = D/(B+D)$. The higher the sensitivity and specificity, the more accurate the test. A higher sensitivity test is better suited to eliminate negative results. A higher specificity, on the other hand, more accurately identifies positive results. Therefore, a test with both high sensitivity and specificity is ideal. These values were calculated for all 848 miRNA samples, using data from all 167 of the patients.

|  | Samples Positive | Samples Negative |
|---|---|---|
| Test Positive | A (true postive) | B |
| Test Negative | C | D (true negative) |

Table 1: An illustrative example comparing samples versus a test.

The expression levels were read into a matrix $G^*$ of dimensions 848x167. The rows thus represented the randomized miRNAs and columns represented patient samples. The first 73 columns corresponded to cancerous samples, and the next 94 to normal samples. However, the samples were randomized within the cancerous and normal sub-groups. The miRNA expression levels varied between 1 and 39185.92. Furthermore, each miRNA had a different maximum expression level. For use in each of the cancer diagnostic algorithms to follow, it proves useful to normalise these data by dividing the expression levels for each miRNA (row) by the maximum expression value of that miRNA. That is, matrix elements $G^*_{ij}$ were divided by $G^m_i$ defined as $G^m_i = \max_j[G^*_{ij}]$ for $j = 1,...,167$, to obtain normalized values $G_{ij} = G^*_{ij}/G^m_i$ lying in the interval [0,1]. The normalised expression values were assembled into a matrix denoted by $G$, also of dimensions 848x167. Training and testing data sets were extracted from $G$. The 73 cancerous patient samples were separated into $m^1_c$ training and $m^2_c$ test samples, and the 94 normal patient samples into $m^1_n$ training and $m^2_n$ test samples. The training data were read into a matrix $G^1$ of dimensions 848x($m^1_c+m^1_n$), whose first $m^1_c$ columns held cancerous samples and next $m^1_n$ columns held normal samples. Similarly, The test data were read into a matrix $G^2$ of dimensions 848x($m^2_c+m^2_n$), whose first $m^2_c$ columns held cancerous samples and next $m^2_n$ columns held normal samples. The algorithm for computing sensitivity and specificity follows.

Consider the $i^{th}$ miRNA. For this miRNA, a threshold value $T_i$ needed to be chosen, such that if this miRNA's normalized expression level for the $j^{th}$ sample satisfied $G^1_{ij} > T_i$ it was considered upregulated, and this miRNA test indicated that the $j^{th}$ sample was cancerous (positive). Conversely, if $G^1_{ij} \leq T_i$ this miRNA test indicated that the $j^{th}$ sample was normal (negative). If the $i^{th}$ miRNA returned $p_i$ positive results when applied to the $m^1_c$ cancerous samples in the training set and $n_i$ negative results when applied to the $m^1_n$ normal samples in the training set, its sensitivity and specificity were given by $s^n_i = p_i/m^1_c$ and $s^p_i = n_i/m^1_n$, respectively. This is evident on reviewing the example presented in the context of Table 1. Note that the sensitivity and specificity satisfy $0 \leq s^n_i \leq 1$ and $0 \leq s^p_i \leq 1$, for the $i^{th}$ miRNA, $i = 1,...,848$.

Clearly, the sensitivity and specificity so computed, $s^n_i$ and $s^p_i$, depend on the corresponding thresholds, $T_i$, for $i = 1,...,848$. To circumvent this subjectivity, each threshold was chosen as that value returning the largest values of $s^n_i + s^p_i$. This ensured that for the $i^{th}$ miRNA a $T_i$ value returning $s^n_i > 0.5$ and $s^p_i > 0.5$ would be preferred over a $T_i$ value returning $s^n_i = 0$ and $s^p_i = 1$ or $s^n_i = 1$ and $s^p_i = 0$, thus favoring simultaneous maximization of $s^n_i$ and $s^p_i$. It is to be noted that choosing a single threshold in the interval, say $T$ satisfying $0 \leq T \leq 1$, to be used with all 848 miRNAs would fail to capture the characteristics of each miRNA as a distinctive test.

With the sensitivity and specificity of each miRNA determined by coding this algorithm in R, all 848 miRNAs were applied to the test data in the matrix $G^2$ to evaluate their effectiveness at identifying cancerous (positive) and normal (negative) samples. See the Results section.

**2.2 A supervised machine learning binary classification algorithm to identify cancer**

Machine Learning algorithms constitute a class of Artificial Intelligence methods that enable the learning of models and decision making with minimal intervention from the scientist [15]. They have proved particularly powerful for complex problems in science and engineering, as well as in marketing, image, text and speech recognition [16], predictions in sports, business and politics [11, 12] and even autonomous driving [17]. They are very effective on problems with large datasets, which may otherwise be difficult to gain insight to. Machine learning algorithms include regression fits (developing a mathematical model to fit data) or classification methods (grouping data). Since, in this study, it is of interest to separate samples into cancerous and normal classes, binary classification methods are appropriate. The training dataset, with samples already labelled as cancerous and normal, can be used to "train" the binary classification algorithm, and the approach can be further described as "supervised learning". Trained, supervised machine learning classification algorithms can be applied to test datasets to make predictions. Machine learning algorithms are increasingly being used in computational biology. For some examples see Brown et al. [18] and Ben-Hur et al. [19]. Here, a supervised machine learning binary classification algorithm is trained and applied to prediction of cancer in the test dataset. It was developed by following the Massively Open Online Course "Machine Learning" offered by Prof. Andrew Ng of Stanford University on the Coursera platform [20].

In machine learning, each sample, whether in the training or test dataset is described by a fixed number of features collected into a feature vector, say $X^j$ for each sample $j$. For the dataset obtained from GEO, the individual miRNA expression levels are the features for each sample. Therefore, there were 167 samples in the matrix $G$, with $m^1_c + m^1_n$ samples in the matrix $G^1$, and 167 - $m^1_c$ - $m^1_n$ samples in the matrix $G^2$. The feature vector $X^j$ of each sample $j$ was an 848 dimensional vector with as many features. It is also common in machine learning algorithms to normalize the feature vectors so that the values $X^j_i$ of the $i^{th}$ feature lie in the interval [0,1] as $j$ runs through the samples. This was already satisfied, since for instance, with the training set, we have $X^j_i = G^1_{ij}$, for $i = 1,...,848$ and $j = 1,...,m^1_c + m^1_n$

The supervised machine learning binary classification algorithm is based on a decision function

$$h(X^j) = \frac{1}{1+\exp(-\Theta^T X^j)} \qquad (1)$$

that takes on values between 0 and 1. Here, $\Theta$ is a vector of coefficients that weight each feature. The idea behind training the machine learning algorithm is to design a method to optimize the values of the components of $\Theta$ so that $h(X^j)$ is as close to the actual data for the $j^{th}$ sample. The data for the training set define a function $Y(X^j)$ that takes on the value 1 if the $j^{th}$ sample is cancerous, and 0 if it is normal. The vector $\Theta$ is optimized by minimizing a cost function, $J(\Theta)$, which is defined for the training set as

$$J(\Theta) = - \sum_{j=1}^{m_c^1+m_n^1} \left( Y(X^j) \log h(X^j) + (1 - Y(X^j)) \log(1 - h(X^j)) \right) + \frac{1}{2}\lambda |\Theta|^2 \qquad (2)$$

For a different dataset the sum would run over the number of samples. The term within the summation is called a logistic cost function. The second term is included to regularize the training method, thus preventing it from resulting in very large values of the components of $\Theta$, which can lead to over-fitting the training data. If the training data are over-fit, the algorithm performs very well on these data, but may provide poor predictions on actual test data. The method for optimizing $\Theta$ begins with choosing initially randomized values of the 848 components of $\Theta$ in the interval [-1,1]. This initial guess of randomized $\Theta$ values is improved as the algorithm is trained and converges to optimized values of $\Theta$. This is done by an elementary gradient descent method that updates $\Theta$ iteratively by combining the following formulas for the $(k+1)^{th}$ iterate from the $k^{th}$ iterate.

$$\Theta^{k+1} = \Theta^k - \alpha \frac{dJ(\Theta)}{d\Theta}\bigg|_{\Theta_k}, \qquad \frac{dJ(\Theta)}{d\Theta} = \sum_{j=1}^{m_c^1+m_n^1} \left( h(X^j) - Y(X^j) \right) X^j + \lambda \Theta \qquad (3a,b)$$

until the cost function $J(\Theta)$ is sufficiently small. A suitably small learning rate $\alpha$ must first be chosen. With $\Theta$ thus optimized, this trained machine learning algorithm can be applied to test samples, $\mathbf{Z}^j$ which are the columns of $\mathbf{G}^2$, and give the probability that test sample $j$ (the $j^{th}$ column of $\mathbf{G}^2$) is cancerous. This algorithm was also coded into an R script shown running in Figure 2.

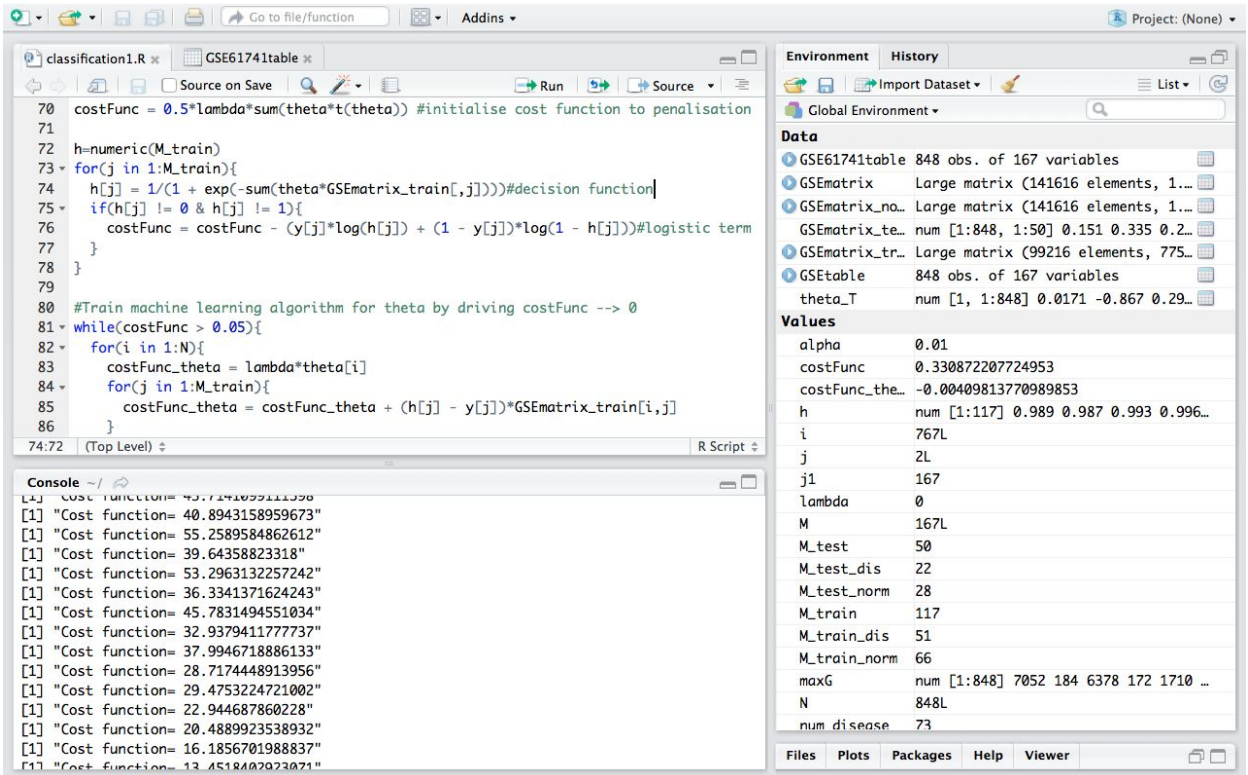

**Figure 2.** The supervised, binary classification machine learning algorithm running in R. The training step is shown with decreasing cost function.

## 3. Results

The results of training the Threshold Algorithm and the supervised machine learning binary classification algorithm are discussed below, as well as their performance on the test data.

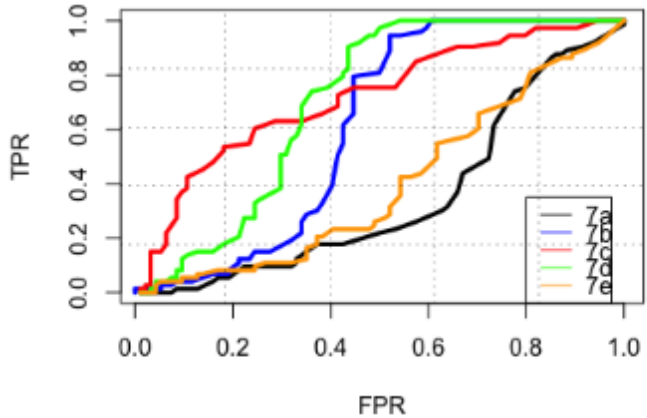

**Figure 3.** The RoC curves for the five miRNAs hsa-let-7a--hsa-let-7e using the entire data set.

**3.1 Calibration of the Sensitivity/Specificity Algorithm for determining and testing results**

The Receiver Operating Characteristic curve (ROC) shows the variation in the true positive rate, or TPR (also the sensitivity) *versus* the false positive rate or FPR (also equal to 1 - specificity) as the threshold is varied over the range of the values in the test [21]. The ROC curve for five of the miRNAs: hsa-let-7a--hsa-let-7e are shown in Fig 1. The entire dataset of 167 samples was used for these ROCs.

The sensitivity/specificity algorithm computes $T_i$ values individually for each of the 848 miRNAs using the training data set. The values of $T_i$ for the miRNAs with the 5 highest sensitivities and specificities varied between 0 and 0.99. The results of applying the 5 highest sensitivity and 5 highest specificity miRNAs to the cancerous and normal test samples were collected for $m^1_c$ and $m^1_n$ chosen so that the training set included 10%, 20%,...,70% of the samples in the cancerous and normal data sets. The miRNAs with the five highest sensitivities and specificities were chosen for each training set, and compared against the corresponding test set (the dataset minus training set) to determine the effectiveness of the corresponding miRNAs as tests for detecting cancer, and for ruling it out. Tables 2 and 3 present the results for the case where the training set contained 10% of samples in the data set, and Tables 4 and 5 present results for the case where the training set contained 70% of the samples in the dataset

**Table 2.** Five highest sensitivity miRNAs with $m^1_c = 7$ and $m^1_n = 9$ (10% of data set).

| miRNA | hsa-let-7a | hsa-let-7b | hsa-let-7c | hsa-let-7d | hsa-let-7e |
|---|---|---|---|---|---|
| TPR | 1.00 | 1.00 | 0.97 | 0.98 | 1.00 |
| FPR | 1.00 | 0.94 | 0.92 | 0.90 | 1.00 |

**Table 3.** Five highest specificity miRNAs with $m^1_c = 7$ and $m^1_n = 9$ (10% of data set).

| miRNA# | hsa-let-7d* | hsa-let-7f-2* | hsa-miR-106a* | hsa-miR-10b | hsa-miR-10b* |
|---|---|---|---|---|---|
| TNR | 0.69 | 0.39 | 0.60 | 0.67 | 0.92 |
| FNR | 0.12 | 0.52 | 0.70 | 0.56 | 0.84 |

**Table 4.** Five highest sensitivity miRNAs with $m^1_c = 51$ and $m^1_n = 66$ (70% of data set).

| miRNA | hsa-let-7i* | hsa-miR-103 | hsa-miR-106a | hsa-miR-106b | hsa-miR-107 |
|---|---|---|---|---|---|
| TPR | 1.00 | 1.00 | 0.95 | 1.00 | 0.95 |
| FPR | 1.00 | 1.00 | 0.96 | 1.00 | 1.00 |

**Table 5.** Five highest specificity miRNAs with $m^1_c = 51$ and $m^1_n = 66$ (70% of data set).

| miRNA | hsa-miR-1 | hsa-miR-1271 | hsa-miR-136 | hsa-miR-146b-3p | hsa-miR-152 |
|---|---|---|---|---|---|
| TNR | 0.86 | 0.96 | 1.00 | 0.86 | 1.00 |
| FNR | 0.91 | 1.00 | 1.00 | 1.00 | 1.00 |

The data in Tables 2-5 show that the highest sensitivity miRNAs obtained from each training set have high true positive rates (TPRs) and the corresponding highest specificity miRNAs have high true negative rates (TNRs). However the high sensitivity miRNAs also have high false positive rates (FPRs), while the high specificity miRNAs have high false negative rates (FNRs). Recall that a high sensitivity test is meant to be useful to identify normal (negative) samples (Section 2.1). However, the high FPRs of the high sensitivity miRNAs in Tables 2 and 4 compromises their effectiveness as diagnostic tests for identifying normal (negative) samples. Similarly, a high specificity test is meant to be useful to confirm diseased (positive) samples (Section 2.1). However the high FNRs of the high specificity miRNAs compromises their effectiveness as diagnostic tests for confirming diseased (positive samples). In this context only miRNA hsa-let-7d* in Table 3 (Column 1) is a somewhat viable test for confirming disease because it has a high TNR and low FNR. No other miRNA is comparable as a test to confirm disease, and no miRNA meets the bar for ruling out normal (negative) cases.

Tables 2--5 also show that increasing the training set does not markedly improve the reliability of the miRNAs as diagnostic tests using sensitivity and specificity. The results in Table 5 are poorer than in Table 3 even though the data in Table 5 is seven times as large as in Table 3. This is surprising; a well-designed test should improve in its predictivity with larger training sets.

### 3.2 Training the machine learning algorithm and test results

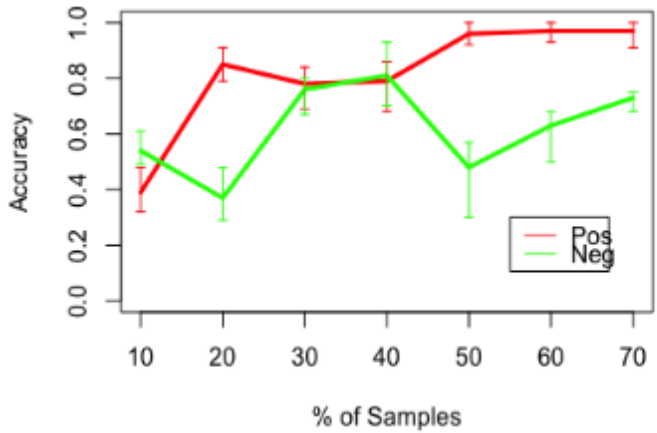

**Figure 4.** Variation in the machine algorithm's average accuracy at identifying positive and negative samples from the test set as the training data set increases to encompass 10% to 70% of the samples. Error bars show maximum and minimum accuracy over 5 training runs.

The supervised binary classification machine learning algorithm was trained with learning rate $\alpha = 0.01$ and regularization parameter $\lambda = 0$. Recall that $\lambda = 0$ allows the algorithm to fit the training data very closely, and even over fit it. However, the test results were not adversely affected by this choice as shown below. After some numerical experiments, cost function $J \leq 0.05$ was selected as a suitable threshold for training the algorithm. If overfitting resulted from $\lambda = 0$ the tolerance on the cost function would also need to be higher than the low value of 0.05 used here. A result $h(X^j) > 0.5$ was taken as a positive result, and $h(Z^j) \leq 0.5$ was taken as a negative result on a test sample with feature vector $Z^j$ (miRNA expression values). Figure 2 shows the variation in the accuracy of the machine learning algorithm, evaluated against the test set, as the training set is expanded to include from 10% to 70% of the samples in steps of 10%. This is the same range used for the study of sensitivity and specificity in Section 3.1. The two curves represent the machine learning algorithm's average accuracy at identifying cancerous/positive (red) and normal/negative (green) samples from the test set. The averages are obtained over 5 training runs of the machine learning algorithm for each training dataset. The error bars show maximum and minimum accuracy attained over the corresponding training set. It is noteworthy that the accuracy at identifying positive samples is well above 0.75 for training sets that consist of 20%, or more, of the samples. The accuracy is 0.97 for training sets containing 60% and 70% of the samples. The accuracy at identifying negative samples is not as high, but again shows an increasing trend with values over 0.5 for 4 of the 7 training sets, and 0.73 for the largest training set. This somewhat lower accuracy than for positive samples may continue to improve with larger training sets. This approach was not pursued, because the statistics from the test sets would become sparse as the total number of samples is fixed at 167 in the GSE61741 data set.

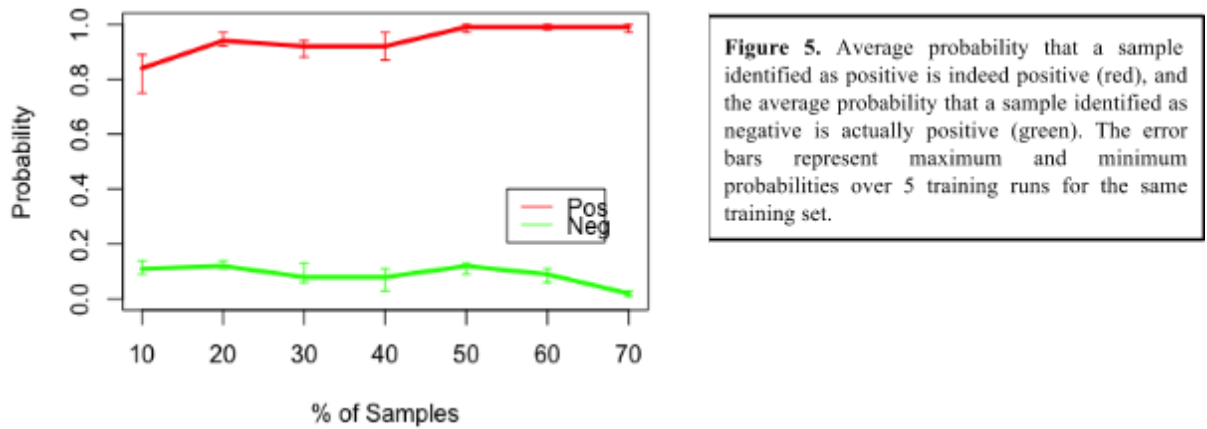

**Figure 5.** Average probability that a sample identified as positive is indeed positive (red), and the average probability that a sample identified as negative is actually positive (green). The error bars represent maximum and minimum probabilities over 5 training runs for the same training set.

The red curve in Figure 5 shows the average probability of a positive, or cancerous outcome for the samples identified as being positive or cancerous *versus* the training data set size. This probability reaches close to 1 for the larger training sets. The green curve is, conversely, the average probability of a positive, or cancerous outcome for the samples identified as being negative/normal *versus* the training data set size. This probability reaches a low of 0.03 for the larger training sets. These are important measures of the confidence in the predictions (accuracy) presented in Figure 4, and can prove useful in actual clinical diagnoses. A higher probability that a sample identified as being cancerous is indeed cancerous is an important measure to gauge reliability of the test. The same can be said if there is a lower probability that a sample identified as being normal is actually cancerous. Averages were computed over 5 training runs, each starting from randomized initial values of the machine learning parameters, $\Theta$. Error bars are maximum and minimum values for the corresponding training set. The uniformity of results indicates the robustness of the algorithm, and the tightening error bounds with larger training sets (Figures 4 and 5) are indications of the improvement of predictions with larger training datasets. This improvement in predictive accuracy is expected of a good machine learning algorithm.

## 4. Discussion

The advent of genomics has brought many advances in medicine and biotechnology. The most exciting of these are in the diagnosis and treatment of diseases. Given the complexities of cancer [22] it is not surprising that there is considerable interest in applying genomics to its diagnosis and treatment. This study has investigated the role of miRNAs in cancer diagnosis. Large scale miRNA sequencing [23] has made it possible to use 100s of them as distinct tests for diseased and normal states. However, this also presents a "big data" problem , where the volume of information available can overwhelm the ability to use it in a meaningful manner. This is where computational biology, and specifically, computational genomics, becomes relevant. Statistical, data-driven algorithms can be used to analyze the genomic data and make predictions that mere observation of 100s of miRNA array tests cannot discern. Here, the sensitivity and specificity, which are traditional  statistical measures, have been compared against a modern machine learning algorithm for identifying cancerous and normal samples using miRNA expression data.

      The data set in this study included a large number of miRNAs (848), and a moderate number of samples (167), which were separated into training and test sets for the calibration and

validation of the algorithms. It is notable (Tables 2--5) that the high sensitivity miRNAs had high false positive rates (FPRs) on the test data, compromising their effectiveness as tests for identifying normal samples (Tables 2 and 4). Similarly, the high specificity miRNAs had high false negative rates (FNRs) on the test data, compromising their effectiveness as tests for identifying cancerous samples (Tables 3 and 5).

These results, while seeming to undermine sensitivity and specificity as statistical measures in computational genomics-based diagnostics, should be approached with caution. They have been obtained with a single data set, which while not small, may have unknown biases in it. It is therefore necessary to apply the sensitivity/specificity tests to more data sets of greater population diversity before reaching definitive conclusions on their effectiveness.

The notable features of the supervised binary classification machine learning algorithm in Section 3.2 are its high accuracies at identifying both cancerous (positive) and normal (negative) samples in the test sets (Figure 4), and the accompanying probabilities of correctness of these predictions (Figure 5). Importantly, this algorithm, trained on the identical datasets as the sensitivity/specificity tests, shows clear improvements in its accuracy at predicting both cancerous and normal samples, and in the probabilities of correctness of these predictions, with increasing training set size. This trend is expected of high performance prediction methods.

As with the sensitivity and specificity measures, the machine learning algorithm also could be trained and tested against larger and more diverse data sets. Given the encouraging preliminary results, it may be expected that it will similarly perform well.

**4.1 Placing this work in context**

The present work adds to the growing literature on computational genomics, especially on machine learning approaches for miRNA-based diagnosis of cancer. A small collection of these studies is available in Refs 24-29. The studies in Refs 24, 25, 26 and 29 used a different machine learning algorithm: a support vector machine (SVM). The others were the iterative signature algorithm [27] and a Naïve Bayesian classifier [28]. These machine learning algorithms were applied to study breast, urothelial, colorectal and lung cancers. The sample sizes ranged from 36 to 141 [24, 25, 29], and the numbers of miRNAs from 12 [25, 27] to 866 [29].

Relative to the above works, the present study used a training sample sizes that were comparable: 16 to 117, with corresponding test datasets ranging from 151 down to 50, respectively. The high probabilities of correctness of the results on the test sets have already been discussed. The feature vector size, i.e., the number of miRNAs in the present study compares with the largest in the above studies [29], which also was a GSE dataset. It is likely that the high accuracies gained are related to this large feature vector. The binary classification algorithm here maybe considered somewhat more elementary than an SVM, but its simplicity could be an advantage in encouraging its use by others, which is a goal of future work (Section 4.2).

**4.2 Future work**

To extend this study, both algorithms will be challenged with larger and more diverse data sets from the GEO repository. This study has potential for broad impact in medicine, because both algorithms developed and tested here can be applied to many medical conditions that allow a binary classification into diseased and normal states. They therefore serve as quite general models for computational genomics-based diagnostic tests. The R codes for both algorithms are well-commented, allowing even a beginning user of R to easily modify and apply them to other data sets, or even to studies of different conditions. A more complex condition involving multiple classes would require a different, multi-class classification machine learning algorithm, which is a planned future extension of this work. This could, for instance, be trained against an miRNA array data set with small cell lung cancer, non-small cell lung cancer and lung carcinoid tumor samples and then against sub-types of other cancers and non-cancer diseases.

Another direction of investigation will extend the machine learning algorithms and codes to select the most effective feature vectors (miRNAs). This could also suggest which miRNAs have similarity of action. All R scripts for both algorithms are available on the GitHub repository https://github.com/neerja-g/machine-learning-miRNA under the LGPL 3.0 license.

Ultimately, the goal is to bring such computational genomics into the clinical setting, where decisions could be reached with greater confidence, given the very quantitative measures of accuracy and probability that the machine learning algorithms provide. Even further, there exists the possibility of alerting a currently healthy individual to the odds of developing a condition in the future. This of course, would also require longer term studies on the change in performance of the algorithms in terms of outcomes, accuracies, and probabilities over many years as the data from the same set of patients changes with their state of health.

**Acknowledgement**

I thank Dr. Inhan Lee of miRcore for introducing me to computational genomics and for her guidance on the initial miRNA expression data analysis from GEO.

**References**

1. American Cancer Society. www.cancer.org

2. P. Baldi and A. Long. A Bayesian framework for the analysis of microarray expression data: regularized **t**-test and statistical inferences of gene changes. *Bioinformatics* **17**(6), 509-519, 2001.

3. J. Lu, G. Getz, E. A. Miska, E. Alvarez-Saavedra, J. Lamb, D. Peck, A. Sweet-Cordero, B. L. Ebert, R. H. Maki, A. A. Ferrando, J. R. Downing, T. Jacks, H. R. Horvitz and T. R. Golub. MicroRNA expression profiles classify human cancers. *Nature*, **435(**9), 834-839, 2005.

4. L. J. Chin, E. Ratner, S. Leng, R. Zhai, S. Nallur, I. Babar, R-U. Muller, E. Straka, L. Su, E. A. Burki, R. E. Crowell, R. Patel, T. Kulkarni, R. Homer, D. Zelterman, K. K. Kidd, Y. Zhu, D. C. Christiani, S. A. Belinsky, F. J. Slack, and J. B. Weidhaas. A SNP in a let-7 microRNA complementary site in the KRAS 3 untranslated region increases non–small cell lung cancer risk. *Cancer Research*, **68**(20), 8535-8540, 2008.

5. G. Rabinowits, C. Gerçel-Taylor, J. M. Day, D. D. Taylor and G. H. Kloecker. Exosomal microRNA: A diagnostic marker for lung cancer. *Clinical Lung Cancer*, **10**(1), 42-46, 2009.

6. M. Boeri, C. Verria, D. Contea, L. Roza, P. Modenab, F. Facchinettia, E. CalabROC, C. M. CROCed, U. Pastorinoc, and G. Sozzia. MicroRNA signatures in tissues and plasma predict development and prognosis of computed tomography detected lung cancer. *PROCeedings of the US National Academy of Science*, **108**(9), 3713–3718, 2011.

7. F. Bianchi, F. Nicassio, M. Marzi, E. Belloni, V. Dall'Olio, L. Bernard, G. Pelosi, P. Maisonneuve6, G. Veronesi and P. P. Di Fiore. A serum circulating miRNA diagnostic test to identify asymptomatic high-risk individuals with early stage lung cancer. *EMBO Molecular Medicine*, **3**, 495–503, 2011.

8. Gene Expression Omnibus. http://www.ncbi.nlm.nih.gov/geo/

9. B. Scholkopf, K. Tsuda and J-P. Vert. Kernel Methods in Computational Biology. MIT Press. 2004.

10. R. Gentleman, W. Huber, V. J. Carey, R. A. Irizarry and S. Dudoit. *Bioinformatics and Computational Biology Solutions using R and Bioconductor*. Springer Pub. 2005.

11. D. Barber. Bayesian Reasoning and Machine Learning. Cambridge University Press. 2012.

12. P. Kulkarni. Reinforcement and Systematic Machine Learning for Decision Making. Wiley. IEEE Press. 2012.

13. R.V. Davuluri, I. Grosse and M.Q. Zhang. Computational identification of promoters and first exons in the human genome. *Nature Genetics*, **29**, 412–417, 2001.

14. E. Mathe, M. Olivier, S. Kato, C. Ishioka, P. Hainaut and S. V. Tavtigian. Computational approaches for predicting the biological effect of p53 missense mutations: a comparison of three sequence analysis based methods. *Nucleic Acids Research*, **34**(5), 1317–1325, 2006.

15. P. Baldi and S. Brunak. Bioinformatics. *The Machine Learning Approach*. MIT Press. 2001.

16. P. Perner (ed). Machine Learning and Data Mining in Pattern Recognition. 6th International Conference MLDM, 2009, Leipzig, Germany, July 23-25, 2009.

17. H. Cheng. Autonomous Intelligent Vehicles. Theory, Algorithms and Implementation. Springer-Verlag, London. 2011.

18. M. P. S. Brown, W. N. Grundy, D. Lin, N. Cristianini, C. W. Sugnet, T. S. Furey, M. Ares, Jr., and D. Haussler. Knowledge-based analysis of microarray gene expression data by using support vector machines. *PROCeedings of the US National Academy of Science*, **97**(1), 262–267, 2000.

19. A. Ben-Hur, C. S. Ong, S. Sonnenburg, B. Schölkopf and G. Rätsch**.** Support Vector Machines and Kernels for Computational Biology. *PLoS Computational Biology*, **4**(10), e1000173, 2008.

20. A. Ng. *Machine Learning*. Coursera, https://www.coursera.org/learn/machine-learning

21. T. Sing, O. Sander, N. Beerenwinkel and T. Lengauer. ROCR: visualizing classifier performance in R. *Bioinformatics Applications Note* **21**(20), 3940–3941, 2005.

22. R. A. Weinberg. *The Biology of Cancer*. Garland Science, 2007.

23. D. Witten, R. Tibshirani, S. G. Gu, A. Fire and W-O. Lui. Ultra-high throughput sequencing-based small RNA discovery and discrete statistical biomarker analysis in a collection of cervical tumours and matched controls. *BMC Biology* **8**, 58, 2010.

24. G. Bertoli, C. Cava and I. Castiglioni. MicroRNAs as Biomarkers for Diagnosis, Prognosis and Theranostics in Prostate Cancer. *International Journal of Molecular Sciences* **17**, 421, 2016.

25. N. Sapre, G. Macintyre, M. Clarkson, H. Naeem, M. Cmero, A. Kowalczyk, P. D. Anderson, A. J. Costello, N. M. Corcoran and C. M. Hovens. A urinary microRNA signature can predict the presence of bladder urothelial carcinoma in patients undergoing surveillance. *British Journal of Cancer* **114**, 454-462, 2016.

26. R. Kothandan and S. Biswas. Identifying microRNAs involved in cancer pathway using support vector machines. *Computational Biology and Chemistry* **55**, 31-36, 2015.

27. A. Fiannaca, M. La Rosa, L. La Paglia, R. Rizzo and A. Urso. Analysis of miRNA expression profiles in breast cancer using biclustering. *BMC Bioinformatics* **16**, S7, 2015.

28. R. Amirkhah, A. Farazmand, S. K. Gupta, H. Ahmadi, O. Wolkenhauer and Ulf Schmitz. Naïve Bayes classifier predicts functional microRNA target interactions in colorectal cancer. *Molecular BioSystems* **11**, 2126-2134, 2015.

29. S. Paul and P. Maji. $\mu$HEM for identification of differentially expressed miRNAs using hypercuboid equivalence partition matrix. *BMC Bioinformatics* **14**, 266, 2013.
22